# HIGH-PERFORMANCE SOLIDLY MOUNTED BIDIMENSIONAL MODE RESONATORS (S2MRs) OPERATING AROUND 16 GHz

*Luca Spagnuolo, Luca Colombo, Kapil Saha, Gabriel Giribaldi, Pietro Simeoni, Matteo Rinaldi*
*Institute for NanoSystems Innovation (NanoSI), Northeastern University, Boston, MA, USA*

## ABSTRACT

This paper reports on Solidly-Mounted Bidimensional Mode Resonators (S2MRs) utilizing 30% Scandium-doped Aluminum Nitride (ScAlN) on Silicon Carbide (SiC), operating near 16 GHz. Experimental results show mechanical quality factors ($Q_m$) up to 380, electromechanical coupling coefficients ($k_t^2$) of 4%, and an overall Figure of Merit ($FOM = Q \cdot k_t^2$) exceeding 15. Additionally, $Q_{Bode}$ calculation is reported along with an analysis of the piezoelectric energy confinement coefficient $\eta$ showing the impact of thickness-to-wavelength ratio ($h/\lambda$) on the acoustic wave confinement. Finally, the devices demonstrate power handling capabilities greater than 20 dBm while achieving close impedance matching to 50 Ω. These features make them strong candidates for commercial, military, and harsh-environment applications like satellite communications (SATCOM) and Active Electronically Scanned Arrays (AESA).

## INTRODUCTION

Emerging communication paradigms, such as 5G and 6G, are creating demand for compact, efficient, and scalable RF components capable of operating in the Ku band and mmWave frequencies [1][2]. Satellite communications (SATCOM) and military applications, such as Active Electronically Scanned Arrays (AESA) [3], could see significant advantages from technologies that offer monolithic integration, robust power handling, and reliable operation under high-temperature conditions. In traditional devices insertion loss (IL) and fractional bandwidth (FBW) decreases significantly at frequencies beyond 5 GHz. This performance degradation is largely due to fabrication limitations, such as the difficulty of achieving ultra-thin films to scale the frequency.

XBAW [4] and periodically poled piezoelectric film (P3F) lithium niobate (LiNbO3) [5] address this limitation by enabling high-frequency operation without relying on extreme miniaturization of film thickness. In this context, Solidly Mounted Bidimensional Mode Resonators (S2MRs) operating at 16 GHz are presented in this paper. This class of devices utilizes an optimized slow-on-fast Sezawa mode, in a thin film of 30% Scandium Aluminum Nitride (ScAlN) on Silicon Carbide (SiC) to attain high mechanical quality factors, high electromechanical coupling, and excellent power handling. Their reduced fabrication complexity and their compatibility with monolithic integration on high-power electronic platforms like SiC and diamond, make these devices a potential solution for high-performance filtering

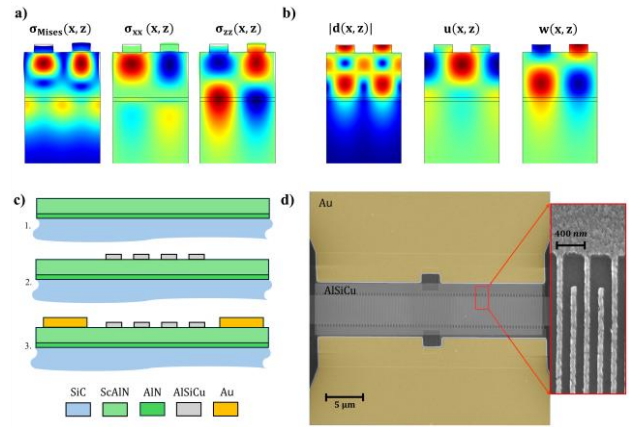

*Figure 1: a) Von Mises equivalent stress; b) displacement magnitude (d); c) micro-fabrication process flow: Top to bottom: 1) ScAlN reactive sputtering on top of an ultra-thin AlN seed layer; 2) Top electrode patterning via electron beam (e-beam) lithography, aluminum-silicon-copper (AlSiCu) thermal evaporation, and lift-off; 3) Pads and interconnects patterning via direct laser lithography; d) SEM picture of final device.*

applications spanning both commercial and military sectors.

## MODELING

S2MRs are a highly optimized Sezawa mode in which a surface acoustic wave is constrained in the piezoelectric layer while solidly mounted on a substrate with high propagation velocity ($v_p$). Similarly to Cross-Sectional Lamé Mode Resonators (CLMRs), the device leverages both the $d_{33}$ and $d_{31}$ piezoelectric coefficients to excite a two-dimensional acoustic standing wave which possess high stress and displacement components in the lateral (x) and thickness (z) directions (Figure 1a, 1b). A critical feature of the design is the acoustic impedance mismatch between the active piezoelectric layer and the substrate. This mismatch effectively confines the acoustic energy within the piezoelectric medium due to the slow-on-fast phase velocity discontinuity and a thickness-to-wavelength ($h/\lambda$) ratio below a critical threshold [6], enabling the propagation of a Sezawa mode with large motional quality factor ($Q_m$) and electromechanical coupling ($k_t^2$).

## FABRICATION

The fabrication process used for the S2MRs is depicted in Figure 1c. First, 230 nm of 30% ScAlN is deposited in-house on a 4-in SiC substrate through reactive sputtering (Evatec CLUSTERLINE® 200 II system). A 20 nm Aluminum Nitride (AlN) seed layer is used as a nucleation layer template to facilitate the growth of a highly crystalline piezoelectric film. Then, Interdigitated Fingers (IDTs) with a width of 100 nm are patterned via e-beam lithography to target the desired frequency of 16 GHz. A 50 nm film of Aluminum Silicon Copper (AlSiCu) is deposited by thermal evaporation, followed by an overnight lift-off in Microposit Remover 1165. Lastly, pads are patterned via maskless laser lithography followed by 250nm e-beam evaporation of gold and lift-off. The final device is shown in Figure 1d.

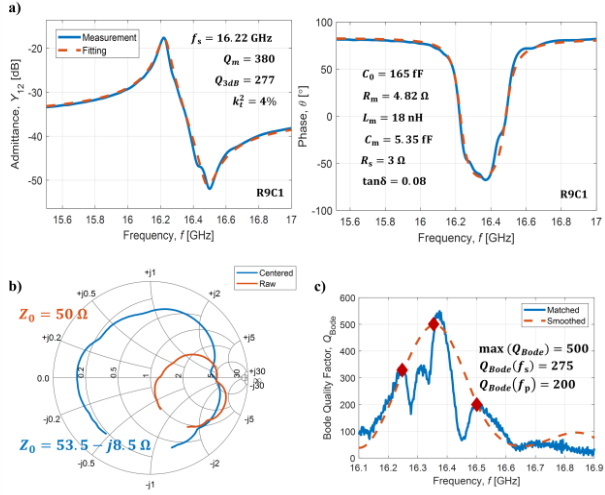

*Figure 2: a) Admittance response (Y12) and phase (θ) of a fabricated S2MR operating around 16 GHz and exhibiting the largest 3-dB quality factor (Q3dB); b) Impedance matching on Smith chart before and after the matching algorithm; c) $Q_{Bode}$ profile and extracted maximum value from smoothing.*

## EXPERIMENTAL RESULTS

The fabricated device is measured with 150 μm GSG probes to extract its frequency response. Scattering (*S*-). parameters are acquired by a Keysight P5008A vector network analyzer (VNA). The measured S-parameters are then converted into admittance (*Y*-) parameters via MATLAB scripts. To extract the equivalent parameters, the admittance ($Y_{12}$) response is fitted to a modified Butterworth-Van Dyke model (mBVD) equivalent circuit model [7]. Figure 2a displays the admittance response of the best device with the highest measured 3-dB quality factor ($Q_{3dB} = 277$), a motional quality factor of $Q_m = 380$, and an electromechanical coupling coefficient $k_t^2$) of 4%. These values yield a figure of merit ($FOM \approx Q_m \cdot k_t^2$) slightly above 15.

Additionally, the extraction of the $Q_{Bode}$ [8] is reported for the same device. The first step involves impedance matching using a Smith chart, ensuring that the response near the resonant frequency is centered on the chart. Once matched, as illustrated in Figure 2b, $Q_{Bode}$ can be calculated using the following Eq. (1):

$$Q_{bode} = \omega \cdot \frac{|S_{11}| group\_delay(S_{11})}{1-|S_{11}|^2} \quad (1)$$

Furthermore, a finite element method (FEM) analysis was performed to investigate the piezoelectric energy confinement coefficient ($\eta$) across various thickness-to-wavelength ratios ($h/\lambda$) [9]. In more detail, $\eta$ quantifies the ratio between the elastic strain energy stored in the piezoelectric material and the total elastic strain energy in the system. It can be estimated using Eq. (2).

$$\eta = \frac{\iiint_{material} W(x,y,z) dxdydz}{\iiint_{tot} W(x,y,z) dxdydz} \quad (2)$$

The results, presented in Figure 3a, demonstrate an increase in the piezoelectric energy confinement coefficient ($\eta_{piezo}$) as the thickness of the piezoelectric layer increases. Conversely, Figure 3b reveals a decrease in the energy confinement within the silicon carbide (SiC) substrate, as reflected in the reduction of the energy confinement coefficient ($\eta_{SiC}$). This can be associated with the Sezawa mode, where, beyond a certain thickness-to-wavelength ratio value of the piezoelectric layer, the energy no longer propagates into the substrate.

Moreover, Figure 3c shows that as the electrode thickness increases, more energy ($\eta_m$) is confined within the electrodes, as indicated by the increase in the electrode energy confinement coefficient ($\eta_m$). This highlights the role of electrode thickness design in energy confinement and the potential for optimizing electrode thickness. As a result, $Q$ and $k_t^2$ can be optimized by adjusting the thickness of various layers (piezoelectric, electrodes and substrate), provided that the intrinsic material quality factors are known [9].

Lastly, the device's power handling capabilities are illustrated in Figure 3d, which shows that the device operates effectively across a power range from 0 dBm to 20 dBm. The legend describes the test procedures, with the black dashed line indicating 0 dBm after a full cycle. The shift in $f_s$ observed at higher power levels is likely a result of device heating. Over this range, the device maintains stable performance without failure, indicating its robustness and suitability for high-power applications.

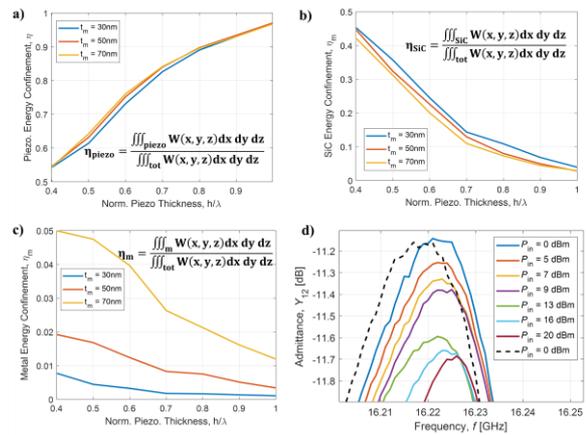

*Figure 3: a) Piezoelectric energy confinement coefficient as a function of thickness to wavelength ratio (h/λ); b) SiC energy confinement coefficient as a function of thickness to wavelength ratio (h/λ); c) Metal energy confinement coefficient as a function of thickness to wavelength ratio (h/λ); d) Power handling test on fabricated S2MR.*

## CONCLUSIONS

In this study, the design, characterization, and performance of Solidly-Mounted Bidimensional Mode Resonators (S2MRs) with a particular focus on their energy confinement, power handling capabilities, and frequency response. Through a combination of finite element method (FEM) analysis and experimental characterization, the influence of key structural parameters on the device's performance is demonstrated.

To highlight the advantages of this design, a performance comparison table (Table 1) is presented benchmarking the fabricated device against state-of-the-art Ku-band resonators. The results demonstrate that the proposed S2MR achieves superior Key Performance Indicators (KPIs), particularly in terms of quality factor and electromechanical coupling compared to other solidly mounted devices and achieves higher power handling capabilities than suspended resonators.

This performance positions S2MRs as a promising candidate for the monolithic integration of nanoacoustic devices with high-power electronics on silicon carbide (SiC), enabling robust solutions for commercial, military, and harsh-environment applications.

*Table 1: Performance comparison of the proposed S2MR with state-of-the-art solidly mounted and suspended resonators operating in the Ku-band. Key metrics include resonant frequency, quality factors ($Q_{3dB}$), electromechanical coupling coefficient ($k_t^2$), and maximum power handling capability.*

| Ref. | $f_s$ [GHz] | $Q_{3dB}$ | $Q_{Bode}$ | $k_t^2$ (%) |
|---|---|---|---|---|
| **This work** | 16.22 | 277 | 500 | 4 |
| [10] | 15.83 | 254 | - | 4.25 |
| [10] | 17.76 | 208 | - | 5.58 |
| [11] | 21.4 | 62 | 66 | 7 |
| [12] | 18.6 | 156 | 210 | 2 |
| [13] | 14.47 | 105 | - | 4 |
| [13] | 16.21 | 55 | - | 5.8 |
| [14] | 14.73 | 155 | 443 | 2.3 |
| [15] | 12.9 | 224 | - | 3.7 |
| [15] | 21.4 | 287 | - | 1.5 |
| [15] | 29.9 | 328 | - | 0.94 |
| [16] | 17.9 | 236.6 | - | 11.8 |

## ACKNOWLEDGMENTS

This work was supported by the Defense Advanced Research Project Agency (DARPA) COFFEE project under Contract HR001122C0088 and developed in collaboration with Raytheon Missiles & Defense (RTX) and Naval Research Lab (NRL). The authors would also like to thank Northeastern University Kostas Cleanroom and Harvard CNS staff.